# Identification of Néel vector orientation in antiferromagnetic domains switched by currents in NiO/Pt thin films

C. Schmitt[1,†], L. Baldrati[1,†], L. Sanchez-Tejerina[2,3], F. Schreiber[1], A. Ross[1,4], M. Filianina[1,4], S. Ding[1,4,5], F. Fuhrmann[1], R. Ramos[6,*], F. Maccherozzi[7], D. Backes[7], M. A. Mawass[8], F. Kronast[8], S. Valencia[8], E. Saitoh[6,9,10,11,12], G. Finocchio[2], M. Kläui[1,4]

[1]*Institute of Physics, Johannes Gutenberg-University Mainz, 55128 Mainz, Germany*

[2]*Department of Mathematical and Computer Sciences, Physical Sciences and Earth Sciences, University of Messina, 98166 Messina, Italy*

[3]*Department of Biomedical, Dental, Morphological and Functional Imaging Sciences, University of Messina, 98125 Messina, Italy*

[4]*Graduate School of Excellence Materials Science in Mainz, 55128 Mainz, Germany*

[5]*State Key Laboratory for Mesoscopic Physics, School of Physics, Peking University, Beijing 100871, China*

[6]*WPI-Advanced Institute for Materials Research, Tohoku University, Sendai 980-8577, Japan*

[7]*Diamond Light Source, Chilton, Didcot, Oxfordshire OX11 0DE, United Kingdom*

[8]*Helmholtz-Zentrum Berlin für Materialien und Energie, Albert-Einstein-Strasse 15, D-12489 Berlin, Germany*

[9]*Institute for Materials Research, Tohoku University, Sendai 980-8577, Japan*

[10]*The Institute of AI and Beyond, The University of Tokyo, Tokyo 113-8656, Japan*

[11]*Center for Spintronics Research Network, Tohoku University, Sendai 980-8577, Japan*

[12]*Department of Applied Physics, The University of Tokyo, Tokyo 113-8656, Japan*

*\*Present address: Center for Research in Biological Chemistry and Molecular Materials (CIQUS), Departamento de Química-Física, Universidade de Santiago de Compostela, Santiago de Compostela 15782, Spain*

† These two authors contributed equally to this manuscript.

ABSTRACT

Understanding the electrical manipulation of the antiferromagnetic order is a crucial aspect to enable the design of antiferromagnetic devices working at THz frequency. Focusing on collinear insulating antiferromagnetic NiO/Pt thin films as a materials platform, we identify the crystallographic orientation of the domains that can be switched by currents and quantify the Néel vector direction changes. We demonstrate electrical switching between different T-domains by current pulses, finding that the Néel vector orientation in these domains is along $[\pm 5\ \pm 5\ 19]$, different compared to the bulk <112> directions. The final state of the in-plane component of the Néel vector $\boldsymbol{n_{ip}}$ after switching by current pulses $\boldsymbol{j}$ along the $[1 \pm 1\ 0]$ directions is $\boldsymbol{n_{ip}} \parallel \boldsymbol{j}$. By comparing the observed Néel vector orientation and the strain in the thin films, assuming that this variation arises solely from magnetoelastic effects, we quantify the order of magnitude of the magnetoelastic coupling coefficient as $b_0 + 2b_1 = 3 \times 10^7$ J/m$^3$. This information is key for the understanding of current-induced switching in antiferromagnets and for the design and use of such devices as active elements in spintronic devices.

INTRODUCTION

Antiferromagnetic materials (AFMs) are promising for spintronic applications, offering several advantages compared to ferromagnets, such as potentially higher switching speeds due to THz resonance frequencies, a higher bit packing density due to the absence of stray fields and increased stability due to the insensitivity to external magnetic fields [1]. The use of AFMs in applications, however, requires an efficient reading and writing of information in defined states of the Néel vector $\boldsymbol{n}$. Recently, electrical switching has been reported for both metallic AFMs [2–4] and insulating AFM/heavy metal bilayers [5–11], however, the underlying mechanism, in particular in the latter case, is under debate [5–8]. The different proposals for the switching mechanism depend on the type of domains. So far, it is largely unknown for which direction of the Néel vector and between which domains the electrical switching occurs. Moreover, the switching can be influenced by mechanisms indirectly related to the injected current flow. For example, the proposed thermomagnetoelastic mechanism for switching relies on a torque via inhomogeneous current-induced temperature gradients generating strain in the thin film via magnetoelastic coupling [8,12]. Among the antiferromagnetic materials where switching has been shown, the most studied to date is NiO, as it provides an ideal platform for spintronic devices. However, the dependence of the magnetoelastic coupling constant on strain in NiO thin films is known only from simulations [13,14], so an experimental estimation is necessary for theoretical models to simulate which effect dominates the switching mechanism.

NiO is a collinear antiferromagnet with a bulk Néel temperature of $T_N = 523$ K [15]. In bulk single crystals, below $T_N$, NiO contracts along the $\langle 111 \rangle$ directions, forming so called T-domains, leading to the spins being confined to four equivalent ferromagnetic {111} planes, coupled antiferromagnetically. Within each of these planes, the spins can orient along one of the three <112> directions (S-domains), leading to a total of 12 possible domain orientations [15–17]. In thin films of NiO grown on MgO(001), it has previously been discussed that strain induces a preferential out-of-plane orientation of ***n*** with respect to the sample plane [18,19]. On the other hand, many reports of the switching of ***n*** between different states [5,7,18] have used spin Hall magnetoresistance (SMR) measurements, which are sensitive to the in-plane components of **n**, and thus indicate the presence of an in-plane component of the Néel vector. It was conjectured that the structure of the magnetic domains in NiO thin films grown on MgO resembles the [±1 ±1 2] of the bulk NiO, where only the domains with large $n_z$, so with large components out of the sample plane, are energetically favorable [18,20,21]. However, a precise determination of the Néel vector direction is missing. The experimental determination of the magnetoelastic coupling constant in particular in thin films has so far not been reported, nor is it clear which types of magnetic domains are involved in the switching of NiO thin films. These pieces of information are key to understand the current-induced switching in antiferromagnets.

In this work, we determine the domain type and the Néel vector orientation of different AFM domains after electrical switching in epitaxial NiO thin films by photoemission electron microscopy (PEEM) employing the x-ray magnetic linear dichroism (XMLD) effect [5,7,22,23]. First, we prepare a state with multiple domains by applying an *in-situ* electric current pulse in the PEEM setup. Second, by analyzing the XMLD signal as a function of the angle between ***n*** and the linear polarization vector, we determine the Néel vector orientation [24], showing that we switch between different T-domains. We further deduce that an electrical pulse along opposite arms of a cross favors a final state with the in-plane component of ***n*** orientated parallel to the current pulse. Finally, we determine the order of magnitude of the magnetoelastic coupling coefficient from the alignment of the Néel vector and, by including the strain applied to the NiO thin films by the substrate, show that this is one order of magnitude larger than predicted from DFT calculations [13,14], suggesting that magnetoelastic effects play an important role in the current-induced switching of antiferromagnetic NiO.

RESULTS AND DISCUSSION

We prepared epitaxial MgO(001)//NiO/Pt(2 nm) samples by reactive magnetron sputtering. After pre-annealing the MgO(001) substrates at 770 °C for 2 hours in vacuum, NiO was deposited from a Ni target at 430 °C and 150 W in an atmosphere of Ar (flow 15 sccm) and $O_2$ (flow 1.5 sccm). The platinum layer was subsequently deposited *in-situ* at room temperature without breaking the vacuum. The epitaxial growth of NiO on MgO (lattice mismatch +0.9%) results in a compressive strain in the out-of-plane direction of the NiO layer (see appendix S1). To be able to apply current pulses, we patterned Hall crosses using optical lithography and subsequent Ar ion etching. The magnetic properties were checked by a polarization-dependent absorption spectrum around the Ni $L_2$ edge at room temperature (Fig. 1(a)). The spectrum shows XMLD, calculated as $I_{LH}$ - $I_{LV}$, indicating antiferromagnetic ordering of the spins [25–27]. To acquire the XMLD images the Ni $L_2$ edge was used and the contrast was calculated as $\frac{I(E_{low})-I(E_{high})}{I(E_{low})+I(E_{high})}$, while we observed no XMCD contrast in images at the $L_3$ edge. For the 10 nm NiO sample in this study, we additionally found a very small magnetic circular dichroism (XMCD) signal. To quantify the relevance of the XMCD signal, we performed SQUID measurements finding a small magnetic remanent moment of ~6 emu/cc which can correspond to a maximum ~1 % Ni phase. In a sample with higher oxygen concentration, we could not detect a measurable ferromagnetic moment stemming from the NiO thin film. As both samples show the same current-induced switching and the same large domains can be imaged by using Kerr microscopy, we can conclude that the presence of a small XMCD signal is not relevant for the domains and their switching as observed here. Fig. 1(b) depicts the device layout and the pulsing scheme used for a MgO(001)//NiO(10 nm)/Pt(2 nm) sample, with a 10 μm Pt cross orientated along the [100] crystallographic axes. The contact pads are not symmetric with respect to the axes of the cross, possibly generating a temperature gradient inclined with respect to [100]. The virgin state of this sample was almost single domain (see Fig. 1(c)), in line with reports of large domains for high quality bulk NiO [17]. Therefore, we first applied a 1 ms-long current pulse with $j = 8.0 \times 10^{11}\ A\ m^{-2}$ along the [100] direction at room temperature *in-situ* in the PEEM microscope, generating a three-domain state (Fig. 1(d)). We studied the XMLD signal by varying the incident in-plane angle of the x-rays $\gamma$ as well as the orientation of the linear polarization $\omega$, both of which are defined in Fig. 1(b).

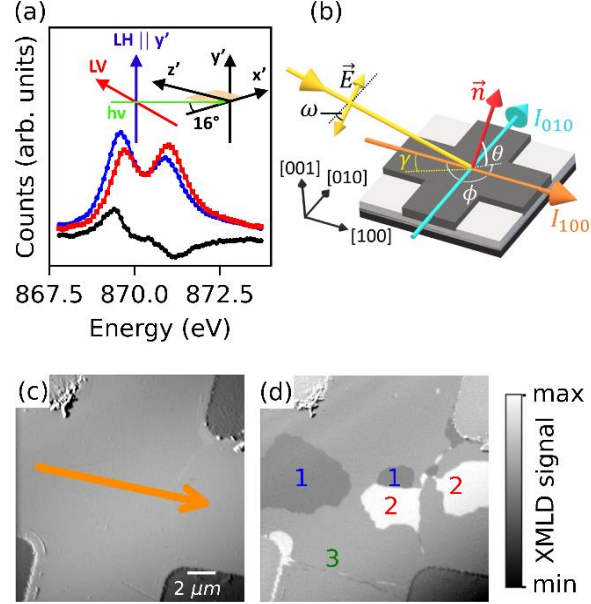

Fig. 1: (a) X-ray absorption spectrum of MgO(001)//NiO(10 nm)/Pt(2) at the Ni $L_2$ edge for linear vertically (LV) and linear horizontally (LH) polarized light. (b) Device layout and pulsing scheme. The Pt cross is oriented along the [100] crystallographic axes. The angles defining the linear polarization vector and the Néel vector are defined with respect to the crystallographic axes. (c) The virgin state of the sample and (d) the three-level contrast after applying $j = 8.0 \times 10^{11}\ A\ m^{-2}$ along the [100] direction for 1 ms.

This causes a different contrast between the domains (Fig. 2(a)-(c)) depending on the direction of **n**, defined by the in-plane angle $\phi$ and the out-of-plane angle $\theta$, in each domain and the projection of **n** on the linear polarization. The angular dependent XMLD signal allows us to determine the Néel vector orientation [24]. Fig. 2(d) and (e) show the XMLD signal as a function of the linear polarization $\omega$ for $\gamma = 0°$ and $\gamma = -45°$, respectively. We take into account the XMLD anisotropy in our tetragonal thin films by a pseudo-spin approximation of the crystalline tetragonal anisotropy. Under this assumption, the intensity at the absorption edge is given by $I = I_0 + I_1 cos^2\alpha + I_2 cos^2\beta$, where $\alpha$ is the angle between the linear polarization and **n** and $\beta$ is the angle between the linear polarization and the crystal field component modelled as a pseudo-spin. We assume that the crystal field is along the out-of-plane direction [001] and independent on the spin axes in the domains, as induced by the out-of-plane strain introduced during the growth (see appendix S2). $I_0, I_1$ and $I_2$ are fitting constants related to the XMLD signal. The XMLD-PEEM images in Fig. 2(a)-(c) reveal that, at $\gamma = -45°$, the contrast between domain 1 (blue) and domain 3 (green) reverses twice upon changing the x-ray linear polarization angle. This is also reflected in Fig. 2(e), where the blue and green curves corresponding to the mentioned domains show two points of intersection at $\omega = 12°$ and $\omega = 75°$. These points of contrast inversion qualitatively and quantitatively determine the relative orientation of the Néel vector in the domains.

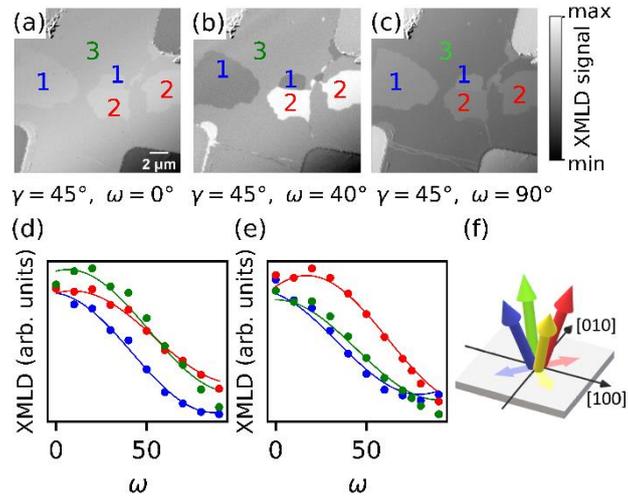

Fig. 2: (a) The three-level contrast for an incident angle of the beam of $\gamma = -45°$ and a polarization $\omega = 0°$. Changing the polarization to (b) $\omega = 40°$ and (c) $\omega = 90°$ changes the relative contrast between the domains. The corresponding XMLD-signal for (d) $\gamma = 27°$ and (e) $\gamma = -45°$ can be fitted including a component stemming from the magnetic signal and an out-of-plane crystal field. (f) The observed domains are T-domains along the four $[\pm 5 \pm 5\ 19]$ directions.

We fitted the different signals and, considering domains compatible with the NiO domain structure, tetragonal distortion, and contrast inversion points, we determined that there are three different domains with ***n*** along the [±5.0 ±5.0 19.0] ± [0.5 0.5 2.0] directions (Fig. 2(f)) revealing a fourfold in-plane symmetry [18,25] and an angle of 20° to the [001] direction. Note that here we do not include the full effect of the anisotropy of the XMLD in NiO as for example proposed in Ref. [26], but with this simple model and the observed angles of the contrast inversion points, we can clearly determine that the out-of-plane Néel vector component is increased compared to the bulk. Based on the symmetry, we identify the domains that can be switched by a current as T-domains, which is a key finding of this work. This is further supported by the comparison with the XLD images at the oxygen-K edge. In Fig. 3 we compare the domain structure of a MgO/NiO(5 nm)/Pt(2 nm) sample, which will be discussed in detail below, obtained by XMLD at the Ni $L_2$ edge and the XLD at the oxygen K edge, acquired with two energies at 530.6 eV and 532 eV. Regardless of the chosen edge we see the same domains, indicating that the observed domain structure can be associated with a domain-dependent crystal field generated by strain and the observed domains are T-domains [27]. The signal to noise ratio in the case of the O-K edge, however, is lower compared to the measurements at the Ni $L_2$ edge (Fig 3(c)).

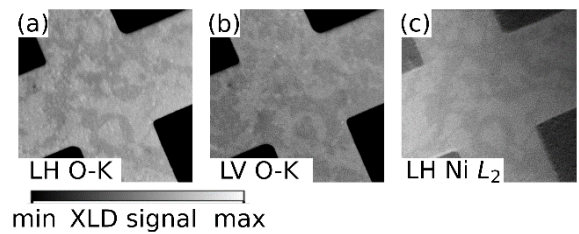

Fig. 3: XMLD images taken at the O-K edge for an incident angle of the beam of $\gamma = 27°$ and (a) linear horizontally and (b) linear vertically polarized light. (c) XMLD image taken on the same sample device at the Ni $L_2$ edge for $\gamma = 27°$ and linear horizontally polarized light, showing the same domain structure.

As already reported, for each T-domain, the S-domain with the largest out-of-plane component is favored [18,21], but, compared to the bulk, the out-of-plane Néel vector component is larger, likely due to the lattice expansion in the in-plane crystal direction. Based on the geometry, one expects the fourth T-domain (in-plane angle $\phi = 135°$) to be present as well, but we did not observe it in the investigated sample area. Note that, with a Kerr microscope-based technique at normal incidence, one can see only two contrast levels out of these four domains [12,21,25]. To investigate electrical switching in these samples, we alternated 1 ms-long current pulses between the [100] and [010] directions. The switching threshold along [100] is observed for a current density of $j = 7.5 \times 10^{11}\ A\ m^{-2}$. A following orthogonal pulse (along [010]), with a slightly lower current density of $j = 7.0 \times 10^{11}\ A\ m^{-2}$, shows that small regions are switched in all three domains.

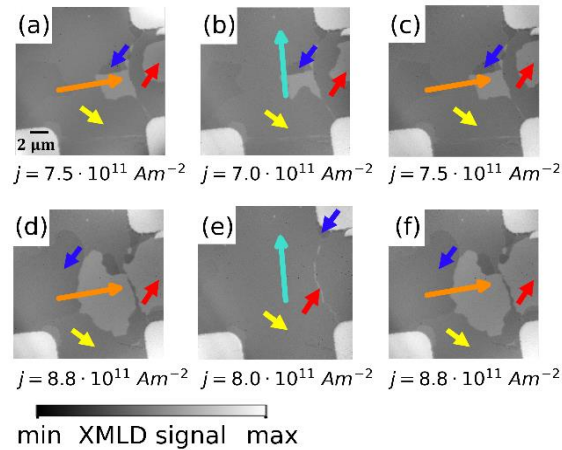

Fig. 4: The application of current pulses at the current threshold alternating between [100] and [010]. (a)-(c) reveals reversible switching of small regions of all three domains. (d) Increasing the current density leads to increased switched regions after a [100] pulse. (e) A perpendicular pulse creates a mono-domain state showing saturation, switching back to the three-level state after a pulse along [100] (f). All images were taken with linearly vertical polarized x-rays and in all images a background intensity was subtracted.

The difference in the current densities necessary to achieve switching is caused by varying resistances along the two arms leading to different heating. Another pulse along [100] fully reverses the switching (Fig. 4(a)-(c)). Upon increasing the current density, the size of the switched regions increases, reaching saturation at $j = 8.8 \times 10^{11} \, A \, m^{-2}$ (Fig. 4(d)) along [100]. A perpendicular current pulse creates a quasi mono-domain state (Fig. 4(e)), similar to the virgin state of the sample, favoring the domain with $\boldsymbol{n}$ orientated along [$\bar{5}$ 5 19]. Another pulse along the [100] direction restores the three-domain state (Fig. 4(f)), showing reproducible current-induced electrical switching for large parts of the sample.

Next, we check if these Néel vector directions are present in other samples. We patterned a MgO(001)//NiO(5 nm)/Pt(2 nm) sample with a 5 µm Hall cross rotated by 45° and symmetric contact pads with respect to the crystallographic axes, see Fig. 5(a). The cross is aligned along [110], such that current pulses can be applied parallel or perpendicular to the in-plane spin direction. This allows us to determine whether a final state of the in plane Néel vector component $\boldsymbol{n}_{ip}$ parallel or perpendicular to the applied pulses is favored. Compared to the previous sample, this sample with thinner NiO layer shows smaller but still switchable domains (Fig. 5(c)). We again see a three-level contrast, which we assign to three T-domains with spin directions [$\pm 5$ $\pm 5$ 19] using an analogous procedure as before. The in-plane projection of $\boldsymbol{n}$ for the three domains, as well as the two possible current pulse directions are shown in Fig. 5(b). Note that domain no. 3 (green), where $\phi$ = -45°, is very small. We now observe the T-domain with the Néel vector orientation that was not present in the previous sample (yellow). Applying alternating current pulses along [110] and [$\bar{1}$10] shows reversible switching (Fig. 5(c)-(e), additional switching cycles and devices are shown in the appendix S3), but likely due to higher pinning, the switched regions are of smaller size compared to the previous NiO(10 nm)/Pt sample (Fig. 4).

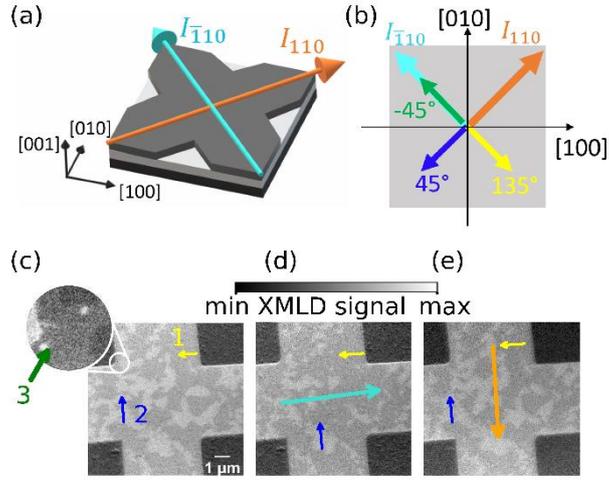

Fig. 5: (a) Device layout and pulsing scheme of the 5 nm NiO film with the Pt cross oriented 45° to the crystallographic axes. (b) Observed in-plane component of Néel vector in the three domains of the NiO(5 nm)/Pt sample and current pulse direction. (c) Domain structure of the virgin state for linear horizontally polarized light. The inset shows a small area of the sample at a different incident angle of the x-rays and a different polarization angle ($\omega = 45°$, $\gamma = 0°$), revealing a third small domain (green). (d) and (e) show the domain structure after 1 ms writing pulses of $j = 1.0 \times 10^{12}$ A m$^{-2}$ in two different directions as indicated by the arrows.

We increased the current density in this sample until electrical breakdown of the device. This occurred before we observed complete switching, in contrast to the thicker 10 nm sample, where we could switch almost the entire device area. Since thinner samples should switch more easily than thicker ones, while here we see the opposite trend, we conclude that this particular sample presents higher pinning, which inhibits full switching. Comparing the domain structure before and after applying the electrical pulses, and knowing the Néel vector orientation in each domain shows that the current induced switching favors a parallel alignment of the in-plane Néel vector component $\boldsymbol{n}_{ip}$ and current pulse direction ($\boldsymbol{n}_{ip} \parallel \boldsymbol{j}$), as some of the authors previously reported by electrical SMR measurements [7], which are, however, prone to spurious signals [28]. This sign is consistent with the thermomagnetoelastic switching mechanism recently reported in similar NiO/Pt samples [12].

In order to extract the quantity that can be used in simulations to understand the switching mechanism, we next estimate the order of magnitude of the magnetoelastic coupling coefficient in NiO. We do this by combining our findings on the Néel vector orientation in the NiO thin films with the strain, induced by the cube-on-cube epitaxial growth that we checked by reciprocal space mapping (see appendix S1). It is generally accepted that the exchange interaction in NiO is along <111> (T-domain) and the dipolar interactions keep the spins in the {111} planes (along e.g. the <112> directions) [15]. Since the dipolar interaction is not significantly affected by strain [29] and as we observe a small change of the spin orientation compared to the bulk, namely an angle of 20° to the [001] direction instead of 35° for the bulk <112> directions, it is reasonable to assume that the small tetragonal distortion along [001], due to the epitaxial growth, contributes to the magnetic anisotropy via the magnetoelastic coupling and favors *n* oriented along the compressed axis [29]. As NiO has a large magnetostriction [30], we thus consider only this effect and neglect other crystalline effects and changes in other interactions, allowing us to estimate the order of magnitude of the magnetoelastic coupling. We can express the magnetoelastic energy as $\varepsilon_{me} = K_{ij}^{ME} m_i^\alpha m_j^\alpha$ [13], where $K_{ij}^{ME}$ is the magnetoelastic constant coupling two spins. According to Ref. [13] $K_{ij}^{ME}$ can be expressed as a function of the strain tensor in NiO. The approximation that the off-diagonal elements are zero leads to:

$$K_{11}^{ME} = (b_0 + 2b_1)e_{11}$$
$$K_{22}^{ME} = (b_0 + 2b_1)e_{22} \quad (1)$$
$$K_{33}^{ME} = (b_0 + 2b_1)e_{33}$$

Where $b_0$ and $b_1$ are the components of the magnetoelastic tensor for NiO, $e_{11}, e_{22}$ are the in-plane strain components and $e_{33}$ is the out-of-plane strain. As shown in appendix S1, we measured $e_{11} = e_{22} = 8.6 \times 10^{-3} \pm 1.2 \times 10^{-3}$, and $e_{33} = -7.1 \times 10^{-4} \pm 9.5 \times 10^{-4}$ by x-ray diffraction. Combining this with the observed Néel vector orientation of $[\pm 5 \pm 5\ 19]$ and setting the bulk magnetocrystalline anisotropy $K_\alpha^{ME} = 0.25$ MJ m$^{-3}$ [30,33], we used micromagnetic simulations to estimate the sum of the magnetoelastic coefficients required to have the experimentally observed equilibrium position of $\bm{n}$ to be $b_0 + 2b_1 = 3 \times 10^7$ J m$^{-3}$ (See appendix S4). This value is an order of magnitude larger than reported in works based on DFT calculations [13,14]. The sign of the coefficients must also be negative, so a compressive (negative) strain enhances the out-of-plane component of the Néel vector. This indicates that the effects due to magnetoelastic coupling can be larger than what was deduced before by DFT calculations, suggesting that thermomagnetoelastic switching can be stronger than spin-orbit torques. This estimation, based on experimental results, can be used for future models and to understand quantitatively the role thermomagnetoelastic effects play in the switching mechanism of antiferromagnets.

CONCLUSIONS

To conclude, we have observed that the electrical switching of NiO/Pt thin films occurs between different T-domains, demonstrating that the switching process involves magnetoelastic effects and that T-domains can be switched relatively easily, as shown for the bulk [17]. We determined that the Néel vector is oriented along the $[\pm 5 \pm 5\ 19]$ directions, canted by 15° towards the out-of-plane direction compared to the bulk [112]. This effect results from the substrate-induced strain with a magnetoelastic coupling coefficient of value $3 \times 10^7$ J m$^{-3}$. Finally, we determined the final state of the in-plane component of the Néel vector to be parallel to the applied electrical pulse ($\bm{n}_{ip} \parallel \bm{j}$) in the center of the device in the presence of pulses along [110], consistent with recent reports on thermomagnetoelastic switching [12]. By knowing the antiferromagnetic domain structure, magnetoelastic coupling coefficient and the final state after switching, one can compare different switching mechanisms, especially those based on thermomagnetoelastic effects and spin-orbit torques.

APPENDIX

**S.1 Strain determination**

The lattice constant of our NiO thin films was determined by x-ray diffraction (XRD) and reciprocal space mapping measurements, using a Bruker D8 Discover high-resolution diffractometer, with $Cu\ K_\alpha$ radiation of wavelength equal to 0.15406 nm. In Fig. S1(a), we show the XRD 2θ-ω scans of the MgO(001)//NiO(10 nm)/Pt(2 nm) sample. The peak position corresponds to the NiO(002) peak, thus, indicating that the NiO orientation is (001) as the MgO substrate. The out-of-plane lattice constant is determined to be $c = 4.173 \pm 0.004$ Å, very similar to the bulk NiO lattice constant of $c = 4.176$ Å. The small lattice mismatch results in an out-of-plane compressive strain of $e_{33} = -7.1 \times 10^{-4} \pm 9.5 \times 10^{-4}$ compared to the bulk value. The large uncertainty is due to the small value extracted for the $e_{33}$ coefficient and due to the presence of the XRD MgO (002) peak in the XRD scan which overlaps with the NiO (002) peak, due to the similar lattice constants. In Fig. S1(b),(c), we show the symmetric and antisymmetric reciprocal space mapping (RSM) at the 002 and 113 diffraction peaks of a MgO//NiO(90 nm)/Pt(2 nm) sample, respectively. The NiO and MgO peak positions are aligned along the same $h$ value, indicating epitaxial growth and a tetragonal distortion. The in-plane lattice parameter of the NiO is $4.212 \pm 0.005$ Å, in line with the MgO lattice parameter. Compared to the NiO bulk value this gives rise to an in-plane strain of $e_{11} = e_{22} = 8.6 \times 10^{-3} \pm 1.2 \times 10^{-3}$.

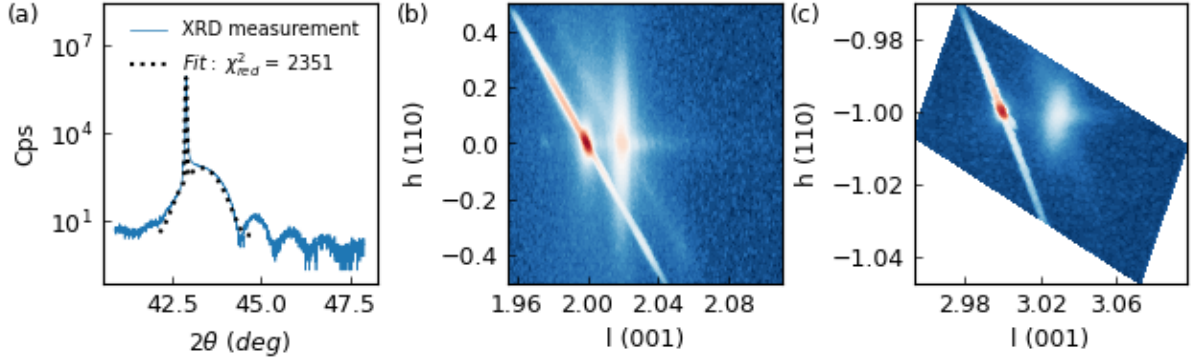

Fig. S1: (a) 2θ-ω XRD scans of the NiO(10 nm)/Pt(2 nm) sample. (b) Symmetric reciprocal space mapping data around the MgO(002) peak. (c) Antisymmetric RSM around the MgO(113) peak, showing the same *h*-value for NiO and MgO.

**S.2 Crystal field component**

In order to explain the trend of the XMLD data, we include a crystal field component as illustrated in Fig. S2. As an example, we consider the MgO(001)//NiO(10 nm)/Pt(2 nm) data for $\gamma = -45°$, included in Fig. 2 (e) of the main text. Simulating the XMLD signal without taking into account a crystal field component leads to the correct simulation of the points of contrast inversion between the different domains and the right relative contrast, but the overall trend of the signal cannot be reproduced (Fig. S2 (a)). However, when we include a crystal field component along the [001] direction, we can leave the angle $\omega$ at which the crossing occurs unchanged and obtain the correct experimentally observed trend. In this way, the acquired data, shown as dashed lines in Fig. S2 (b) can be reproduced. We explain this crystal field component along [001] by the strain induced by the epitaxial growth of NiO on MgO(001) (Fig. S1(c)). The lattice mismatch generates a tetragonal distortion in the NiO thin film and a tetragonal symmetry with compressive strain along the [001] axis resulting in an out-of-plane crystal field. The crystal field component of the XMLD signal caused by this strain is then independent of the spin direction in the domains, as for all of them the angle between the spin axis and the crystal field is equivalent.

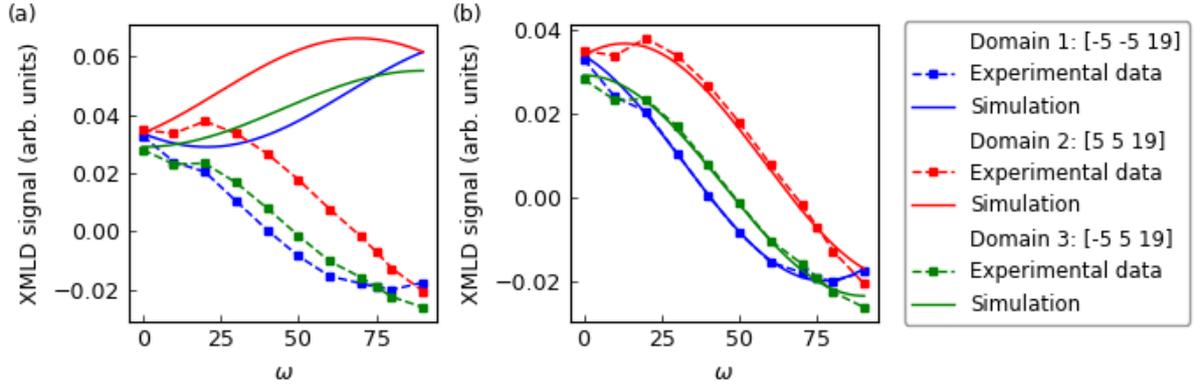

Fig. S2: (a) Simulation of the XMLD signal for the three different domains shown in Fig. 2 of the main text, when considering only a magnetic contribution to the signal. Note that the crossing points are reproduced, but not the overall trend. (b) By including a crystal field component, we obtain good agreement between the simulation and the XMLD data (dashed lines).

**S.3 Switching reproducibility**

To demonstrate the reproducibility of the electrical switching, we include in Fig. S3 additional switching cycles for the NiO(5 nm)/Pt(2) sample shown in Fig. 5(c)-(e). Applying alternating orthogonal pulses switches reliably between two domain configurations. Moreover, we probed the switching in another piece of the NiO(10 nm)/Pt(2) sample imaged in Fig. 2, this time comprising a Hall cross rotated by 45° with respect to the crystallographic axes [100] and [010]. This sample shows larger domains and we performed a complete Néel vector analysis at a spot few hundred microns away from the Hall cross. By comparing this to the contrast levels in the Hall cross, we can identify the domain types and Néel order orientation within the Hall cross. This then allows us to determine that the switching sign when applying an electrical current pulse in a straight configuration is the same for NiO 5 nm and 10 nm, as we show in Fig. S4 (final state with in-plane Néel order direction parallel to the applied pulse).

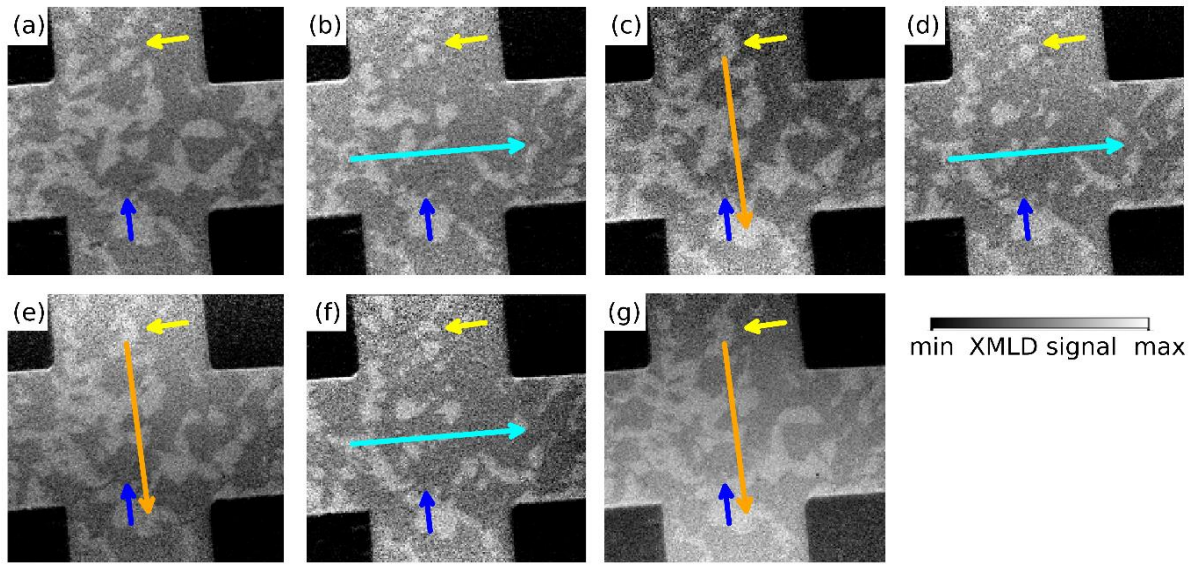

Fig. S3: Applying current pulses in two orthogonal directions shows reproducible switching favoring the in-plane component of the Néel vector to be aligned parallel to the previously applied current pulse. All images were taken with linear horizontally polarized x-rays under an azimuthal angle of $\gamma = 45°$.

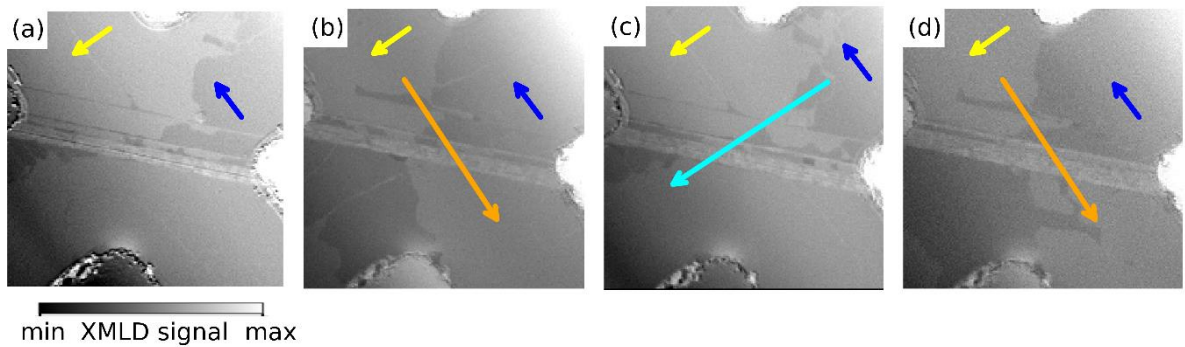

Fig. S4: In a NiO(5 nm)/Pt(2) sample comprising larger domains one can see that the electric switching favors the in-plane component of the Néel vector to be aligned parallel to the applied pulse. Note that here we show images taken for linear vertically polarized x-rays. The approximately horizontal contrast at the centre is topographic contrast of non-magnetic origin.

## S.4 Micromagnetic simulations

The AFM state of the sample can be described by two sublattice magnetizations strongly coupled by the exchange interaction. Each of the sublattices verifies the Landau-Lifshitz-Gilbert equation [33–35]

$$\begin{cases} \frac{d\mathbf{m}_1}{dt} = -\gamma_0 \mathbf{m}_1 \times \mathbf{H}_{eff,1} + \alpha \mathbf{m}_1 \times \frac{d\mathbf{m}_1}{dt} \\ \frac{d\mathbf{m}_2}{dt} = -\gamma_0 \mathbf{m}_2 \times \mathbf{H}_{eff,2} + \alpha \mathbf{m}_2 \times \frac{d\mathbf{m}_2}{dt} \end{cases} \quad (S1)$$

where $\gamma_0$ is the gyromagnetic ratio, $\alpha$ the Gilbert damping coefficient and

$$\mathbf{H}_{eff,i} = -\frac{1}{\mu_0 M_s} \frac{\delta \varepsilon}{\delta \mathbf{m}_i} \quad (S2)$$

the effective field for the i-th sublattice, $M_s$ the sublattice saturation magnetization, $\mu_0$ the vacuum magnetic permeability and $\varepsilon$ the energy density of the system. The energy density includes the contributions from the exchange interaction, the anisotropy field and the magnetoelastic interaction. The expression for the exchange interaction can be found in previous works [34,35]. Hence, we focus on the last two terms. The anisotropy energy can be written as

$$\varepsilon_a(\mathbf{m}_\alpha) = K_a \left(1 - a_p(\mathbf{m}_\alpha)\right)^2, \quad (S3)$$

$$a_1(\mathbf{m}) = \left(1 - \frac{1}{6}(m_x + m_y - 2m_z)^2\right)\left(1 - \frac{1}{6}(m_x - 2m_y + m_z)^2\right)\left(1 - \frac{1}{6}(2m_x - m_y - m_z)^2\right)$$
$$a_2(\mathbf{m}) = \left(1 - \frac{1}{6}(2m_x - m_y + m_z)^2\right)\left(1 - \frac{1}{6}(m_x + m_y + 2m_z)^2\right)\left(1 - \frac{1}{6}(m_x - 2m_y - m_z)^2\right)$$
$$a_3(\mathbf{m}) = \left(1 - \frac{1}{6}(m_x - m_y - 2m_z)^2\right)\left(1 - \frac{1}{6}(2m_x + m_y - m_z)^2\right)\left(1 - \frac{1}{6}(m_x + 2m_y + m_z)^2\right)$$
$$a_4(\mathbf{m}) = \left(1 - \frac{1}{6}(m_x - m_y + 2m_z)^2\right)\left(1 - \frac{1}{6}(2m_x + m_y + m_z)^2\right)\left(1 - \frac{1}{6}(m_x + 2m_y - m_z)^2\right)$$
(S4)

Where $a_p$ allows us to choose between the four possible easy planes. The magnetoelastic contribution reads [13]

$$\varepsilon_{me} = K_{ij}^{ME} m_{\alpha,i} m_{\alpha,j}, \quad (S5)$$

$$K_{ij}^{ME} = b_{ijkl} e_{kl} = \left(b_0 \delta_{ij} \delta_{kl} + b_1 \left(\delta_{ik} \delta_{jl} + \delta_{ik} \delta_{jl}\right)\right) e_{kl}. \quad (S6)$$

$b_0, b_1$ are the components of the magnetoelastic tensor for NiO, and $e_{kl}$ is the strain tensor. Assuming the off-diagonal elements are 0, $K_{ij}^{ME} = 0$, if $i \neq j$,

$$\begin{aligned} K_{11}^{ME} &= (b_0 + 2b_1)e_{11} \\ K_{22}^{ME} &= (b_0 + 2b_1)e_{22} \\ K_{33}^{ME} &= (b_0 + 2b_1)e_{33} \end{aligned} \quad (S7)$$

Taking into account these three contributions (exchange, anisotropy and magnetoelastic) into the effective field, we study the equilibrium configuration of the Néel vector **n** as a function of the parameter $b_0 + 2b_1$ as shown in Fig. S5. The parameters considered for the simulations are: $A_{11} = 5$ pJ/m, $A_0 = -5$ pJ/m, $M_s = 0.35$ MA/m, $K_a = 0.25$ MJ/m³, $e_{11} = e_{22} = 8.6 \times 10^{-3}$ and $e_{33} = -7.1 \times 10^{-4}$. The direction [5 5 19] corresponds to a $z$ component of the Néel vector $n_z = 0.9372$ which is verified for magnetoelastic coupling coefficients of $b_0 + 2b_1 = 3 \times 10^7$ J m$^{-3}$.

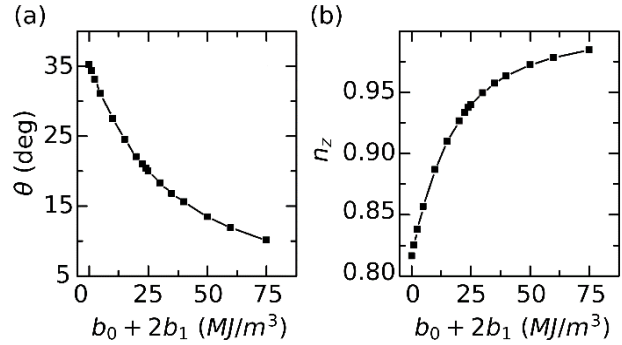

Fig. S5: Equilibrium configuration of the Néel vector $l$ as a function of the magnetoelastic coefficients. (a) Polar angle of the Néel vector at equilibrium. (b) Out-of-plane component of the Néel vector at equilibrium.

ACKNOWLEDGMENTS

L.B acknowledges the European Union's Horizon 2020 research and innovation program under the Marie Skłodowska-Curie grant agreements ARTES number 793159. L.B., A.R., S.D., M.F. and M.K. acknowledge support from the Graduate School of Excellence Materials Science in Mainz (MAINZ) DFG 266, the DAAD (Spintronics network, Project No. 57334897). We acknowledge that this work was funded by the Deutsche Forschungsgemeinschaft (DFG, German Research Foundation) - TRR 173 –268565370 (projects A01 and B02). This publication is based upon work supported by the King Abdullah University of Science and Technology (KAUST) Office of Sponsored Research (OSR) under Award No. OSR-2019-CRG8-4048. This project has received funding from the European Union's Horizon 2020 research and innovation programme under grant agreement No. 863155 (s-Nebula). We acknowledge Diamond Light Source for time on beamline I06 under proposals MM22448 and MM23819-1. This work was also supported by ERATO "Spin Quantum Rectification Project" (Grant No. JPMJER1402) and the Grant-in-Aid for Scientific Research on Innovative Area, "Nano Spin Conversion Science" (Grant No. JP26103005), Grant-in-Aid for Scientific Research (S) (Grant No. JP19H05600), Grant-in-Aid for Scientific Research (C) (Grant No. JP20K05297) from JSPS KAKENHI. R.R. acknowledges support from the European Commission through under H2020-MSCA-RISE grant agreement 734187 (SPICOLOST), the European´s Union 2020 research and innovation program through the Marie Sklodowska Curie Actions grant agreement SPEC-894006 and the Spanish Ministry of Science (RyC2019- 026915I). L.S.T. and G.F. acknowledges financial support from the program "ASSEGNI DI RICERCA 2019" at the University of Messina and PETASPIN Association. This work was supported by the Max Planck Graduate Centre with the Johannes Gutenberg Universität Mainz (MPGC).

REFERENCES

[1]  V. Baltz, A. Manchon, M. Tsoi, T. Moriyama, T. Ono, and Y. Tserkovnyak, Antiferromagnetic spintronics, Rev. Mod. Phys. 90, 015005 (2018).
[2]  P. Wadley, B. Howells, J. elezny, C. Andrews, V. Hills, R. P. Campion, V. Novak, K. Olejnik, F. Maccherozzi, S. S. Dhesi, S. Y. Martin, T. Wagner, J. Wunderlich, F. Freimuth, Y. Mokrousov, J. Kune, J. S. Chauhan, M. J. Grzybowski, A. W. Rushforth, K. W. Edmonds, B. L. Gallagher, and T. Jungwirth, Electrical switching of an antiferromagnet, Science 351, 587 (2016).
[3]  S. Yu. Bodnar, L. Šmejkal, I. Turek, T. Jungwirth, O. Gomonay, J. Sinova, A. A. Sapozhnik, H.-J. Elmers, M. Kläui, and M. Jourdan, Writing and reading antiferromagnetic Mn2Au by Néel spin-orbit torques and large anisotropic magnetoresistance, Nat. Commun. 9, 348 (2018).
[4]  J. Shi, V. Lopez-Dominguez, F. Garesci, C. Wang, H. Almasi, M. Grayson, G. Finocchio, and P. K. Amiri, Electrical manipulation of the magnetic order in antiferromagnetic PtMn pillars, Nat. Electron. 3, 92-98 (2020).
[5]  T. Moriyama, K. Oda, T. Ohkochi, M. Kimata, and T. Ono, Spin torque control of antiferromagnetic moments in NiO, Sci. Rep. 8, 14167 (2018).
[6]  X. Z. Chen, R. Zarzuela, J. Zhang, C. Song, X. F. Zhou, G. Y. Shi, F. Li, H. A. Zhou, W. J. Jiang, F. Pan, and Y. Tserkovnyak, Antidamping-Torque-Induced Switching in Biaxial Antiferromagnetic Insulators, Phys. Rev. Lett. 120, 207204 (2018).
[7]  L. Baldrati, O. Gomonay, A. Ross, M. Filianina, R. Lebrun, R. Ramos, C. Leveille, F. Fuhrmann, T. R. Forrest, F. Maccherozzi, S. Valencia, F. Kronast, E. Saitoh, J. Sinova, and M. Kläui, Mechanism of N\'eel Order Switching in Antiferromagnetic Thin Films Revealed by Magnetotransport and Direct Imaging, Phys. Rev. Lett. 123, 177201 (2019).
[8]  P. Zhang, J. Finley, T. Safi, and L. Liu, Quantitative Study on Current-Induced Effect in an Antiferromagnet Insulator/Pt Bilayer Film, Phys. Rev. Lett. 123, 247206 (2019).
[9]  I. Gray, T. Moriyama, N. Sivadas, G. M. Stiehl, J. T. Heron, R. Need, B. J. Kirby, D. H. Low, K. C. Nowack, D. G. Schlom, D. C. Ralph, T. Ono, and G. D. Fuchs, Spin Seebeck Imaging of Spin-Torque Switching in Antiferromagnetic $\mathrm{Pt}/\mathrm{NiO}$ Heterostructures, Phys. Rev. X 9, 041016 (2019).
[10] Y. Cheng, S. Yu, M. Zhu, J. Hwang, and F. Yang, Electrical Switching of Tristate Antiferromagnetic Néel Order in α−Fe2O3 Epitaxial Films, Phys. Rev. Lett. 124, 027202 (2020).
[11] L. Baldrati, C. Schmitt, O. Gomonay, R. Lebrun, R. Ramos, E. Saitoh, J. Sinova, and M. Kläui, Efficient spin torques in antiferromagnetic CoO/Pt quantified by comparing field- and current- induced switching, Phys. Rev. Lett. 125, 077201 (2020).
[12] H. Meer, F. Schreiber, C. Schmitt, R. Ramos, E. Saitoh, O. Gomonay, J. Sinova, and M. Kläui, Direct imaging of current-induced antiferromagnetic switching revealing a pure thermomagnetoelastic switching mechanism, Nano Lett. 21(1), 114-119 (2021).
[13] P. A. Popov, A. R. Safin, A. Kirilyuk, S. A. Nikitov, I. Lisenkov, V. Tyberkevich, and A. Slavin, Voltage-Controlled Anisotropy and Current-Induced Magnetization Dynamics in Antiferromagnetic-Piezoelectric Layered Heterostructures, Phys. Rev. Appl. 13, 044080 (2020).
[14] T. Nussle, P. Thibaudeau, and S. Nicolis, Coupling magneto-elastic Lagrangians to spin transfer torque sources, J. Magn. Magn. Mater. 469, 633 (2019).
[15] W. L. Roth, Neutron and Optical Studies of Domains in NiO, J. Appl. Phys. 31, 2000 (1960).
[16] G. A. Slack, Crystallography and Domain Walls in Antiferromagnetic NiO Crystals, J. Appl. Phys. 31, 1571 (1960).

[17] W. L. Roth and G. A. Slack, Antiferromagnetic Structure and Domains in Single Crystal NiO, J. Appl. Phys. 31, S352 (1960).
[18] D. Alders, L. H. Tjeng, F. C. Voogt, T. Hibma, G. A. Sawatzky, C. T. Chen, J. Vogel, M. Sacchi, and S. Iacobucci, Temperature and thickness dependence of magnetic moments in NiO epitaxial films, Phys. Rev. B 57, 11623 (1998).
[19] S. Altieri, M. Finazzi, H. H. Hsieh, H.-J. Lin, C. T. Chen, T. Hibma, S. Valeri, and G. A. Sawatzky, Magnetic Dichroism and Spin Structure of Antiferromagnetic NiO(001) Films, Phys. Rev. Lett. 91, 137201 (2003).
[20] S. R. Krishnakumar, M. Liberati, C. Grazioli, M. Veronese, S. Turchini, P. Luches, S. Valeri, and C. Carbone, Magnetic linear dichroism studies of in situ grown NiO thin films, J. Magn. Magn. Mater. 310, 8 (2007).
[21] J. Xu, C. Zhou, M. Jia, D. Shi, C. Liu, H. Chen, G. Chen, G. Zhang, Y. Liang, J. Li, W. Zhang, and Y. Wu, Imaging antiferromagnetic domains in nickel oxide thin films by optical birefringence effect, Phys. Rev. B 100, 134413 (2019).
[22] P. Wadley, S. Reimers, M. J. Grzybowski, C. Andrews, M. Wang, J. S. Chauhan, B. L. Gallagher, R. P. Campion, K. W. Edmonds, S. S. Dhesi, F. Maccherozzi, V. Novak, J. Wunderlich, and T. Jungwirth, Current polarity-dependent manipulation of antiferromagnetic domains, Nat. Nanotechnol. 13, 362 (2018).
[23] S.-W. Cheong, M. Fiebig, W. Wu, L. Chapon, and V. Kiryukhin, Seeing is believing: visualization of antiferromagnetic domains, Npj Quantum Mater. 5, 3 (2020).
[24] S. M. Czekaj, Ferromagnetic and Antiferromagnetic Domain Configurations in Thin Films and Multilayers: Towards a Patterned Exchange Bias System, ETH Zurich, 2007.
[25] F. Schreiber, L. Baldrati, C. Schmitt, R. Ramos, E. Saitoh, R. Lebrun, and M. Kläui, Concurrent magneto-optical imaging and magneto-transport readout of electrical switching of insulating antiferromagnetic thin films, Appl. Phys. Lett. 117, 082401 (2020).
[26] G. van der Laan, N. D. Telling, A. Potenza, S. S. Dhesi, and E. Arenholz, Anisotropic x-ray magnetic linear dichroism and spectromicroscopy of interfacial Co/NiO(001), Phys. Rev. B 83, 064409 (2011).
[27] K. Arai, T. Okuda, A. Tanaka, M. Kotsugi, K. Fukumoto, T. Ohkochi, T. Nakamura, T. Matsushita, T. Muro, M. Oura, Y. Senba, H. Ohashi, A. Kakizaki, C. Mitsumata, and T. Kinoshita, Three-dimensional spin orientation in antiferromagnetic domain walls of NiO studied by x-ray magnetic linear dichroism photoemission electron microscopy, Phys. Rev. B 85, 104418 (2012).
[28] A. Churikova, D. Bono, B. Neltner, A. Wittmann, L. Scipioni, A. Shepard, T. Newhouse-Illige, J. Greer, and G. S. D. Beach, Non-magnetic origin of spin Hall magnetoresistance-like signals in Pt films and epitaxial NiO/Pt bilayers, Appl. Phys. Lett. 116, 022410 (2020).
[29] M. Finazzi and S. Altieri, Magnetic dipolar anisotropy in strained antiferromagnetic films, Phys. Rev. B 68, 054420 (2003).
[30] N. B. Weber, H. Ohldag, H. Gomonaj, and F. U. Hillebrecht, Magnetostrictive Domain Walls in Antiferromagnetic NiO, Phys. Rev. Lett. 91, 237205 (2003).
[31] H. Kondoh and T. Takeda, Observation of Antiferromagnetic Domains in Nickel Oxide, J. Phys. Soc. Jpn. 19, 2041 (1964).
[32] M. Hutchings and E. Samuelsen, Measurement of Spin-Wave Dispersion in NiO by Inelastic Neutron Scattering and Its Relation to Magnetic Properties, Phys Rev B 6, (1972).
[33] H. V. Gomonay and V. M. Loktev, Spin transfer and current-induced switching in antiferromagnets, Phys. Rev. B 81, 144427 (2010).

[34] V. Puliafito, R. Khymyn, M. Carpentieri, B. Azzerboni, V. Tiberkevich, A. Slavin, and G. Finocchio, Micromagnetic modeling of terahertz oscillations in an antiferromagnetic material driven by the spin Hall effect, Phys. Rev. B 99, 024405 (2019).
[35] L. Sánchez-Tejerina, V. Puliafito, P. Khalili Amiri, M. Carpentieri, and G. Finocchio, Dynamics of domain-wall motion driven by spin-orbit torque in antiferromagnets, Phys. Rev. B 101, 014433 (2020).